\documentclass[10pt, conference, letterpaper]{ IEEEtran}
\usepackage[font=footnotesize]{caption}
\IEEEoverridecommandlockouts
\usepackage{cite}
\usepackage{amsmath,amssymb,amsfonts}
\usepackage{algorithm}
\usepackage{algpseudocode}
\usepackage{etoolbox}
\algrenewcommand\algorithmicdo{}
\algrenewtext{EndFor}{\textbf{End}}
\algrenewcommand\algorithmicfor{\textbf{For}}
\usepackage{etoolbox}

\makeatletter
\newcommand*{\algrule}[1][\algorithmicindent]{%
  \makebox[#1][l]{\hspace*{.8em}%
    \smash{\vrule width 1.0pt height 0.9\baselineskip depth 0.27\baselineskip}%
  }%
}
\newcount\ALG@printindent@tempcnta
\def\ALG@printindent{%
  \ifnum\theALG@nested>0
    \ALG@printindent@tempcnta=1
    \loop
      \algrule[\csname ALG@ind@\the\ALG@printindent@tempcnta\endcsname]%
      \advance\ALG@printindent@tempcnta 1
    \ifnum\ALG@printindent@tempcnta<\numexpr\theALG@nested+1\relax
    \repeat
  \fi
}
\patchcmd{\ALG@doentity}{\noindent\hskip\ALG@tlm}{\ALG@printindent}{}{%
  \errmessage{failed to patch algorithmicx}}

\makeatother

\usepackage{graphicx}
\usepackage{textcomp}
\usepackage{xcolor}
\usepackage{stfloats}
\usepackage{multirow}
\usepackage{tabularx}
\usepackage{subcaption}
\usepackage{diagbox}
\usepackage{mathtools}
\usepackage{booktabs}
\usepackage{svg}
\usepackage{etoolbox}
\begin{document}

\title{Relaying Signal When Monitoring Traffic: Double Use of Aerial Vehicles Towards Intelligent Low-Altitude Networking
}

\author{\IEEEauthorblockN{
    Jiahui Liang,~
    Wenlihan Lu,~
    Tianyi Liu,~
    Kang Kang,~
    Guixin Pan,~
    Liuqing Yang,~
    Xinhu Zheng,~
    Shijian Gao
}

}

\maketitle 

\begin{abstract}
In intelligent low-altitude networks, integrating monitoring tasks into communication unmanned aerial vehicles (UAVs) can consume resources and increase handoff latency for communication links. To address this challenge, we propose a strategy that enables a “\textit{double use}” of UAVs, unifying the monitoring and relay handoff functions into a single, efficient process. Our scheme, guided by an integrated sensing and communication framework, coordinates these multi-role UAVs through a proactive handoff network that fuses multi-view sensory data from aerial and ground vehicles. A lightweight vehicle inspection module and a two-stage training procedure are developed to ensure monitoring accuracy and collaborative efficiency. Simulation results demonstrate the effectiveness of this integrated approach: it reduces communication outage probability by nearly 10\% at a 200~\text{Mbps} requirement without compromising monitoring performance and maintains high resilience (86\% achievable rate) even in the absence of multiple UAVs, outperforming traditional ground-based handoff schemes. Our code is available at the https://github.com/Jiahui-L/UAP.

\end{abstract}


\section{Introduction}

In intelligent low-altitude networks, unmanned aerial vehicles (UAVs) are increasingly tasked with dual roles: serving as aerial relays for robust communication and as mobile sensors for applications like traffic monitoring \cite{35, 11, 34}. However, integrating these functions creates a conflict: both conventional handoff and monitoring schemes that collect separate performance metrics for communication and surveillance incur significant latency that grows with the network size \cite{9, 3}. This makes them unsuitable for time-sensitive applications \cite{20}.


This tension motivates the need for a unified approach. Emerging integrated sensing and communication (ISAC) frameworks suggest that sensor data itself can be used to guide communication decisions \cite{10, 37}. Prior work has explored deep learning based handoff prediction using RF sensing \cite{5}. With the growing deployment of advanced sensors, recent studies have investigated how cameras on roadside units (RSUs) can enhance communication performance by blockage prediction \cite{30, 33} as well as proactive handoff \cite{28, 29, 6}. However, above methods exhibit ineffectiveness in low-altitude networks due to the limited visibility of ground cameras. To address this limitation, the cooperation perception based approaches have been proposed. \cite{7} has used the visual data from UAVs and the location data of ground users for blockage prediction, while \cite{8} has introduced a cooperative LiDAR-based framework for RSU handoffs. However, perspective distortion of images and samples sparsity of cloud points inhibit these schemes' ability to recover accurate ground geometry and global low-altitude semantics, which serve as the critical factors for traffic monitoring via UAVs and reliable handoff between UAV relays and RSUs.



To enable a smarter use of UAVs, this work introduces the \underline{U}nified \underline{A}erial \underline{P}erception \underline{Net}work (UAP-Net), a novel multi-agent scheme designed to seamlessly integrate communication handoff and sensing-based monitoring in low-altitude networks. The core innovation of UAP-Net lies in its fusion of vehicle-mounted LiDAR data with UAV-mounted RGB camera feeds. This multi-modal approach simultaneously captures precise ground geometry, critical for predicting line-of-sight (LoS) links and maintaining stable communication, along with rich global semantic information needed for accurate traffic monitoring. Architecturally, UAP-Net is built around a central perception backbone that employs dedicated feature extraction streams to process data from UAVs and vehicles separately. The outputs from these streams are intelligently merged by a multi-view fusion module, which correlates the disparate features to enable proactive and robust link handoff. To efficiently perform the monitoring task without redundant computation, a lightweight traffic inspection head is directly attached to the backbone, reusing UAV's feature representations. Finally, a two-stage training strategy is employed to coordinate these components effectively, and the entire system is designed for distributed execution, ensuring operational efficiency in real-world deployments. Simulation results validate that UAP-Net reduces communication outage probability by nearly 10\% at a 200~\text{Mbps} requirement without any loss in monitoring performance. Furthermore, the system exhibits remarkable resilience, maintaining over 86\% of the achievable rate even in the absence of multiple UAVs, thereby outperforming traditional ground-based schemes and providing a robust solution for time-sensitive low-altitude applications.

\section{System Model and Problem Formulation}
\begin{figure}[t]
\centering
\includegraphics[width=0.85 \columnwidth]{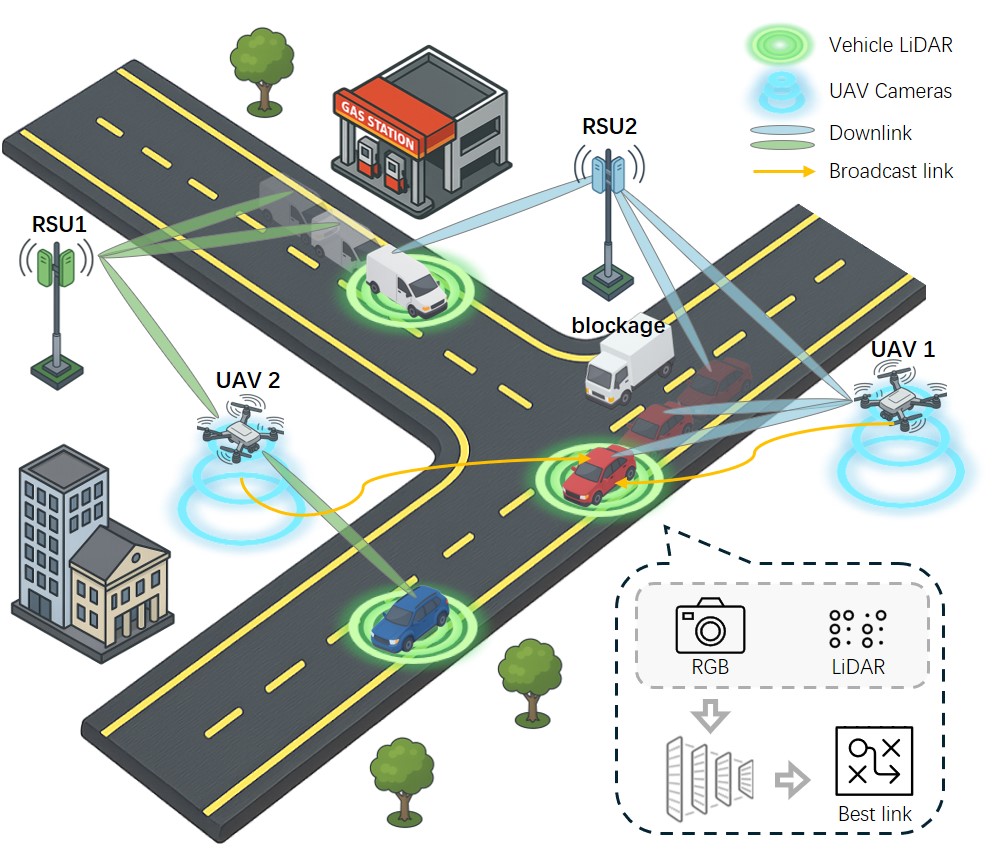}
\caption{An illustration of UAV-assisted air-ground connected networks.}
\label{fig1}
\end{figure}

As illustrated in Fig.~\ref{fig1}, we consider a low-altitude system where ground users perform proactive handoffs between a set of $K$ RSUs and $M$ UAV relays. These UAV relays have the dual responsibility of providing communication links and performing a monitoring task. The system serves $V$ single-antenna ground users. Each RSU is equipped with an $N_r$-element Uniform Linear Array (ULA), while each UAV relay is equipped with an $N_u$-element ULA. We assume the RSUs employ a predefined beamforming codebook. The $k$-th RSU transmits a superposition of normalized data symbols $s_v$ intended for the $v$-th user and its associated set of relays $\mathcal{M}_v$. Therefore, the transmitted signal from the $k$-th RSU can be expressed as:
\vspace{-3pt}
\begin{equation}
\mathbf{x}_k = \sum_{v=1}^{V} \big(\mathbf{w}_{v,k} s_v + \sum_{m=1}^{|\mathcal{M}_v|}\mathbf{w}_{m,k} s_v\big),
\label{eq1}
\end{equation}
where $\mathbf{w}_{v,k}$ and $\mathbf{w}_{m,k}$ represents beamforming vector corresponding to the $v$-th user and the $m$-th relay.

Let $\mathbf{h}_{v,k} \in \mathbb{C}^{N_r \times 1}$, $\mathbf{h}_{m,k} \in \mathbb{C}^{N_r \times N_u}$ and $\mathbf{h}_{v,m} \in \mathbb{C}^{N_u \times 1}$ stand for the channels between RSU-$k$ and user-$v$, RSU-$k$ and UAV-$m$, as well as UAV-$m$ and user-$v$. The UAVs are supposed to adopt the decode-and-forward (DF) mechanism \cite{12} to relay symbols to users. The received signal at user-$v$ can be classified into direct link (DL) or indirect link (IL):
\begin{equation}
\label{eq2}
\begin{aligned}
\textbf{DL}:~& y_{v,k} = \mathbf{h}_{v,k}^{\mathrm H}\mathbf{w}_{v,k} s_v
  + \sum_{i \ne v}\mathbf{h}_{v,k}^{\mathrm H} \mathbf{w}_{i,k} s_i + n_{v,k},\\
\textbf{IL}:~& y_{v,m,k} = \mathbf{h}_{v,m}^{\mathrm H}\mathbf{w}_{v,m} s_v^k
  + \sum_{i \ne v}\mathbf{h}_{v,m}^{\mathrm H} \mathbf{w}_{i,m} s_i^k + n_{v,m}.
\end{aligned}
\end{equation}
Here, $n_{v,k}~n_{v,m}$ are the additive white Gaussian noise with covariance of $\sigma^2$ and $\mathbf{w}_{v,m}$ is the beamforming vector from UAV-$m$ to user-$v$. The symbol $s_{v,k}$ is decoded by UAV-$m$'s received signal $y_{m}=\mathbf{h}_{m,k}^{\text{H}}\mathbf{w}_{m,k} s_v + \sum_{i \ne m}\mathbf{h}_{m,k}^{\text{H}} \mathbf{w}_{i,k} s_i + n_{m,k}$. Thus, the received signals for user $v$ constitute a set $\mathcal{Y}_v = \{\, y_{v,\kappa} \mid \kappa \in \mathcal{K}_v \,\}$, where the index set is shown as $\mathcal{K}_v= \{k\}_{k=1}^K \cup \{(m,k)\}_{m=1}^{|\mathcal{M}_v|}\}_{k=1}^K$ has a cardinality of $|\mathcal{K}_v|=K(|\mathcal{M}_v|+1)$. Accordingly, the set of achievable rates for user-$v$ $\mathcal{R}_v = \{\, R_{v,\kappa} \mid \kappa \in \mathcal{K}_v \,\}$ which is computed as follows:
\vspace{-3pt}
\begin{subequations}
\label{eq3}
\begin{alignat}{3}
\textbf{DL}&:~&& R_{v,k}   & = &~\log_2\!\Big(1 + \frac{|\mathbf{h}_{v,k}^\text{H} \mathbf{w}_{v,k}|^2}{\sum_{i \neq v} | \mathbf{h}_{v,k}^\text{H}\mathbf{w}_{i,k}|^2 + \sigma^2}\Big)\label{eq3a},\\
\textbf{IL}&:~&& R_{v,m,k} & = &~\min\big\{ \tau R_{m,k},~(1-\tau) R_{v,m} \big\}, \label{eq3b}
\end{alignat}
\end{subequations}
where $R_{m,k}$ and $R_{v,m}$ are the achievable rates of the backhaul link (RSU-$k$ to UAV-$m$) and the access link (UAV-$m$ to user-$v$), respectively, both calculated similarly to Eq.~\eqref{eq3a}. The parameter $\tau \in [0,1]$ represents the time fraction allocated for the backhaul transmission. Users in conventional systems must perform exhaustive processes like beam sweeping or channel estimation to collect the rate set ${\mathcal{R}_v}$ and select the optimal link. In our proposed scheme, we bypass these costly steps. Instead, a user directly predicts its best link and executes a proactive handoff based on the cooperative fusion of sensory data from the user itself and the low-altitude UAVs.


Given the above established model, the task of selecting the best link to maximize the sum rate can be formulated as:
\begin{equation}
\label{eq4}
\begin{aligned}
    \max_{\mathbf{\Theta}_g} \quad & \sum_{v=1}^{V} R_{v,\kappa} \\
    \text{s.t.}\quad & \kappa = \mathcal{G}\big(\mathbb{M}_v;\mathbf{\Theta}_g\big),
\end{aligned}
\end{equation}
where $\mathcal{G}(\cdot)$ denotes the neural networks (NN) parameterized by $\mathbf{\Theta}_g$, which is deployed on the vehicle to execute link handoff decisions. This work considers two widely-used sensor modalities: RGB cameras mounted on the UAVs and a LiDAR sensor on the vehicle. The set of informative sensory representations is denoted as $\mathbb{M}_v = \{\mathbf{x}_v^{\text{LiDAR}}, \{\mathbf{x}_m^{\text{RGB}}\}_{m=1}^{|\mathcal{M}_v|} \}$, which includes the LiDAR point cloud from the $v$-th vehicle and the RGB images from UAVs serving it. Additionally, the monitoring task can be formulated as:
\begin{equation}
\label{eq11}
\begin{aligned}
    \min_{\mathbf{\Theta}_d} \quad & \sum_{m=1}^{M} \|\mathbf{d}_m - \hat{\mathbf{d}}_m\|_2^2\\
    \text{s.t.}\quad & \hat{\mathbf{d}}_m = \mathcal{D}\big(\mathbf{x}_m^{\text{RGB}};\mathbf{\Theta}_d\big).
\end{aligned}
\end{equation}
Here, the monitoring output of the $m$-th UAV is defined as $\mathbf{d}_m = [d_1, d_2, \ldots, d_{N_l}]$, where $d_i$ represents the number of vehicles on the $i$-th lane, and $N_l$ is the total number of lanes within the UAV's surveillance coverage. The predicted monitoring vector $\hat{\mathbf{d}}_m$ is obtained by the NN $\mathcal{D}(\cdot)$ parameterized by $\mathbf{\Theta}_d$, which processes the visual representations $\mathbf{x}_m^{\text{RGB}}$ captured by the UAV. The joint optimization objective of this work is to learn the parameters for the handoff network $\mathcal{G}(\cdot)$ and for the monitoring network $\mathcal{D}(\cdot)$.



\section{Unified Aerial Perception Framework}


This section details UAP framework  to enable proactive link handoff during vehicle monitoring. UAP introduces a central backbone for multi-agent feature fusion and handoff prediction with a lightweight inspection head for monitoring. The system is trained through a centralized two-stage procedure and executed in a distributed manner for practical deployment.

\subsection{Unified Aerial Perception Network (UAP-Net)}

Directly fusing raw, heterogeneous data from different agents is ineffective for neural networks. We therefore design separate feature extraction modules for vehicles and UAVs to preprocess data into unified representations containing environmental semantics and ground depth. The following subsections detail their designs.

\begin{figure*}[htbp]
\centering
\includegraphics[width=\textwidth]{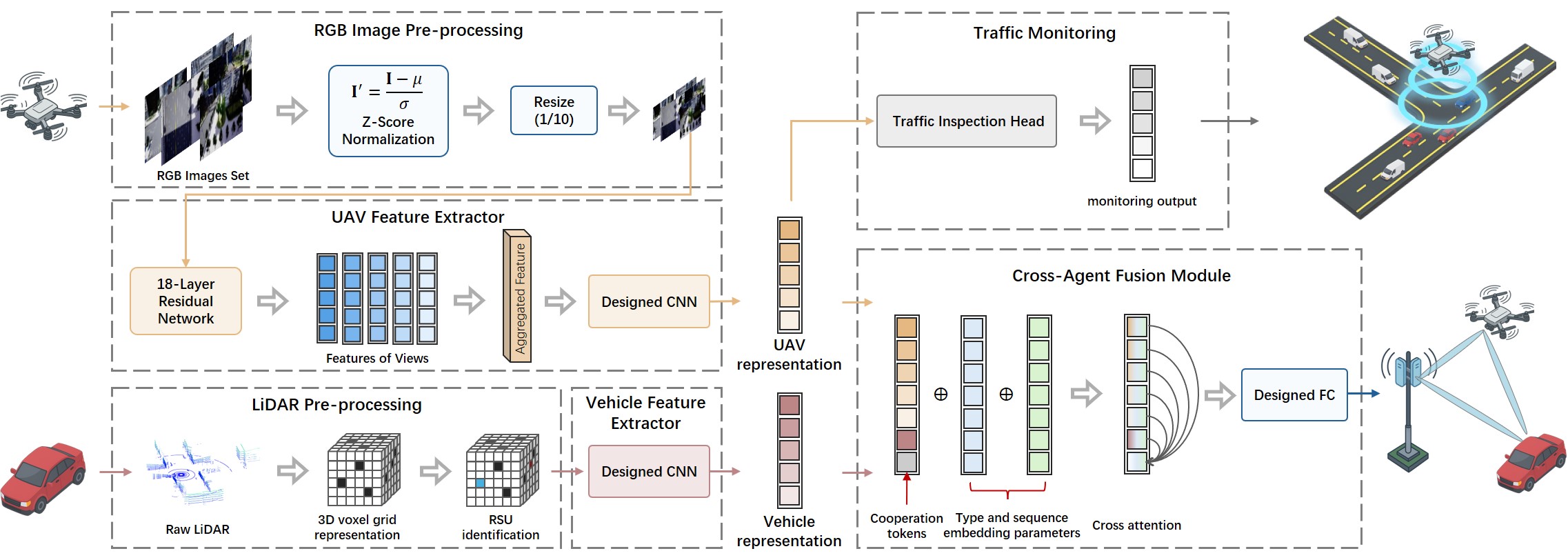}
\caption{An illustration of UAP-Net.}
\label{fig2}
\end{figure*}

\subsubsection{\textit{UAV Feature Extraction Module}}
To capture comprehensive surroundings, each UAV is equipped with cameras on its front, rear, left, right, and downward sides. Let $\mathcal{I}_m = \{\mathbf{I}_{{\rm f},m},\mathbf{I}_{{\rm b},m}, \mathbf{I}_{{\rm l},m}, \mathbf{I}_{{\rm r},m}, \mathbf{I}_{{\rm d},m}\}$ denote the image set from the $m$-th UAV, respectively. First, all RGB images undergo Z-score normalization per channel for stable training:
\begin{equation}
\label{eq5}
\hat{\mathbf{I}}_{{\rm i},m}^{(j)} = \frac{\mathbf{I}_{{\rm i},m}^{(j)} - \mu_m^{(j)}}{\sigma_m^{(j)}}, \quad {\rm i} \in \{{\rm f}, {\rm b}, {\rm l}, {\rm r}, {\rm d}\}, \quad j \in \{1, 2, 3\}
\end{equation}
where $\mu_m^{(j)}$ and $\sigma_m^{(j)}$ are the mean value and the standard deviation of the $j$-th channel. The normalized images are then resized to 10\% of their original resolution to reduce computational load while preserving essential information.

After preprocessing, $\hat{\mathcal{I}}_m$ is fed into the UAV Feature Extractor (UFE), which employs ResNet-18 \cite{13} to efficiently extract visual features while mitigating learning difficulty and gradient vanishing. To match the target output dimension, the original fully connected (FC) layer is replaced with a custom FC layer of length $L_{\text{I}}$ , which defines the visual feature size. The extracted image features for the $m$-th UAV from a set $\mathcal{U}_m = \{\mathbf{f}_{{\rm f},m},\mathbf{f}_{{\rm b},m}, \mathbf{f}_{{\rm l},m}, \mathbf{f}_{{\rm r},m}, \mathbf{f}_{{\rm d},m}\}$ with $\mathbf{f}_{{\rm i},m} \in \mathbb{R}^{1 \times L_{\text{I}}}$. These multi-view features are then aggregated through concatenation:
\begin{equation}
\label{eq6}
\mathbf{f}_{{\rm A},m}= \mathbf{f}_{{\rm f},m} \oplus \mathbf{f}_{{\rm b},m} \oplus \mathbf{f}_{{\rm l},m} \oplus \mathbf{f}_{{\rm r},m} \oplus \mathbf{f}_{{\rm d},m},
\end{equation}
where $\oplus$ denotes the concatenation operation, yielding the aggregated visual feature $\mathbf{f}_{{\rm A},m} \in \mathbb{R}^{5 \times L_{\text{I}}}$. This feature is then processed by a convolutional neural network (CNN) to generate the final informative visual representation $\mathbf{x}_m^{\text{RGB}}\in \mathbb{R}^{L_m}$ for the $m$-th UAV. The complete processing flow of the UAV feature extraction module is illustrated in Fig.~\ref{fig2}.

\subsubsection{\textit{Vehicle Feature Extraction Module}}
LiDAR point clouds collected by the vehicle capture the spatial structure of the communication environment, potentially assisting in modeling the complex relationships among different links through NNs. Since LiDAR point clouds have the property of permutation invariance. Thus conventional CNNs turn out to be unsuitable \cite{8}. To address this issue, we propose a vehicle feature extraction module that first preprocesses LiDAR data from vehicle-$k$ into a structured 3D voxel grid $\mathbf{L}_v$. Using a right-handed coordinate system centered on the vehicle, the LiDAR coverage area is discretized into a ${d_1^L \times d_2^L \times d_3^L}$ voxel grid\footnote{The voxel grid size is selected to balance spatial resolution and computational efficiency within the LiDAR’s effective perception range.}. Each voxel is assigned a value of 1 if it contains at least one point, and 0 otherwise, formally defined as:
\begin{equation}
\label{eq7}
\mathbf{L}_v(i, j, k) =
\begin{cases}
1, & \text{if } N_{v}(i, j, k) > 0 \\
0, & \text{otherwise}
\end{cases},
\end{equation}
where $N_{v}(i,j,k)$ denotes the number of LiDAR points within voxel $(i,j,k)$. To encode RSU semantics, each RSU is assigned a unique identifier $\mathrm{RSU}_k=-k$.  If an RSU is located within a voxel, this identifier replaces the standard binary value. The resulting preprocessed voxel grid $\mathbf{L}_v \in \mathbb{R}^{d_1^L \times d_2^L \times d_3^L}$ is fed into the Vehicle Feature Extractor (VFE) which uses a custom CNN to generate the final informative vehicle representation $\mathbf{x}_v^{\text{LiDAR}} \in \mathbb{R}^{L_v}$.

\subsubsection{\textit{Adaptive Cross-Agent Fusion Module}}

To enable efficient multi-agents cooperation, we design an \underline{A}daptive \underline{C}ross-\underline{A}gent \underline{F}usion (ACAF) module based on the cross-attention mechanism, augmented with type and positional embeddings. This allows the network to adapt to dynamic changes in the number of available UAV views. Specifically, let $\mathbf{x}^{\text{Coop}} \in \mathbb{R}^{L_c}$ denote the cooperation tokens to acts as a collector $\mathbf{Q} = \mathbf{W}_Q \mathbf{x}^{\text{Coop}}$ querying each features of UAVs and vehicles to calculate attention scores and then update aggregated feature $\mathbf{h}^{\text{Coop}} \in \mathbb{R}^{L_c}$: 
\begin{equation}
\label{eq9}
\mathbf{h}^{\text{Coop}} =\sum_i 
\alpha\!\left(\frac{\mathbf{Q} \mathbf{K}_i^{\top}}{\sqrt{L_c}}\right)\mathbf{V}_i.
\end{equation}
where $\mathbf{K}_i = \mathbf{W}_K \mathbf{x}_i$ and $\mathbf{V}_i = \mathbf{W}_V\mathbf{x}_i$ represent the key and value of multi-agent features\footnote{To ensure stable learning, the dimensions of the tokens corresponding to vehicle and UAVs are unified to $L_c$.} with $\mathbf{x}_i \in \{\mathbf{x}_v^{\text{LiDAR}}, \{\mathbf{x}_m^{\text{RGB}}\}_{m=1}^{|\mathcal{M}_v|}\}$. The $\alpha(\cdot)$ is the softmax function for attention scores normalization. According to  Eq.~\eqref{eq9}, the agents information is cross fused by a ever-present $\mathbf{x}^{\text{Coop}}$ and available features, thus ACAF module can adapt various involved views of agents. Additionally, considering the importance of semantics, we introduce type and sequence embedding for agents. For each token, the learnable parameters $\mathbf{e}^{\text{type}}\in \mathbb{R}^{L_c}, \text{type}\in \{\text{LiDAR, RGB, Coop}\}$ and position identifiers $\mathbf{e}^{\text{seq}}\in \mathbb{R}^{L_c}, \text{seq}\in \{1, ..., (2+|\mathcal{M}_v|)\}$ are embedded for type and sequence determination, where $\mathbf{e}^{\text{seq}}$ is computed by:
\begin{equation}
\label{eq8}
\mathbf{e}^{\text{seq}}[j] =
\begin{cases}
\sin\!\left(\frac{\text{seq}}{c^{j/L_c}}\right), & \text{if } j ~\text{is even} \\
\cos\!\left(\frac{\text{seq}}{c^{(j-1)/L_c}}\right), & \text{if } j ~\text{is odd}
\end{cases},
\end{equation}
where $c$ is arbitrary coefficient \cite{15}. The resulting embedded multi-agent features are fused by Eq.~\eqref{eq9} and then fed to a FC to generate the best link.




\subsection{Centralized Training Procedure}

To fulfill traffic monitoring requirements in low-altitude systems, we introduce a lightweight traffic inspection head (TIH) comprising two FC layers, enabling multi-task operation with minimal computational overhead. To prevent negative transfer during multi-task learning, we develop a two-stage centralized training scheme to optimize the above modules, as outlined in Algorithm~\ref{alg1}. Specifically, in stage-I, the UAP-Net is trained as a unified end-to-end NN $\mathcal{G}(\cdot)$ with parameters $\mathbf{\Theta}_g$, consisting of the UFE, VFE, and ACAF, to enable multi-agents cooperative learning across heterogeneous modalities, with the objective of minimizing link prediction error for each vehicle using a cross-entropy loss function:
\begin{equation}
\label{eq10}
\mathcal{L}_{\text{handoff}} = - \sum_{i=1}^{|\mathcal{K}_v|} 
\kappa^{i}\log(\hat{\kappa}^{i}),
\end{equation}
where $\hat{\kappa}^{i}$ is the predicted probability of the $i$-th link. When finishing, stage-II begins to train the TIH $\mathcal{D}(\cdot)$ parameterized by $\mathbf{\Theta}_d$ solely, supported by the UAV feature representations $\hat{\mathbf{x}}_m^{\text{RGB}}$ extracted from the frozen UFE, which prevents distribution shift in the UAP-Net caused by TIH, thereby mitigating negative transfer. According to Eq.~\eqref{eq11}, stage-II adopts mean squared error loss function. Gradient descent is employed to optimize $\mathbf{\Theta}_g$ and $\mathbf{\Theta}_d$.


\subsection{Distributed Execution Scheme}

During the execution phase, the trained UAP-Net and TIH are deployed to their respective agents. The parameters obtained in the centralized training stage are assigned to their respective modules, enabling cooperative handoff and monitoring. On each UAV, the UFE operates independently to extract visual representations $\mathbf{x}_m^{\text{RGB}}$ from real-time RGB images. These features are simultaneously utilized by the TIH for monitoring and are periodically broadcast to vehicles to maintain multi-agent synchronization. On each vehicle, the VFE transforms LiDAR point clouds into structured representations $\mathbf{x}_v^{\text{LiDAR}}$, which are fused with the received visual features $\{\mathbf{x}_m^{\text{RGB}}\}_{m=1}^{|\mathcal{M}_v|}$ from UAVs. The fused heterogeneous features are then processed by the ACAF to initiates proactive link handoff before the current link quality deteriorates. This distributed execution creates a collaborative system where UAVs provide global visual monitoring while vehicles leverage fused multi-view data for stable, low-latency connectivity.


\begin{algorithm}[t]
\caption{Two-stage Centralized Training Procedure}
\label{alg1}
\begin{algorithmic}[1]
\State \textbf{Input:} Data for each vehicle $(\{\hat{\mathcal{I}}_m\}_{m=1}^{|\mathcal{M}_v|}, \mathbf{L}_v, \kappa^*_v)$
\Statex \hspace*{0.96cm} Inspection label for each UAV $\mathbf{d}_m$
\State \textbf{Output:} Trained parameters $\mathbf{\Theta}_g,\mathbf{\Theta}_d$
\State Initialize $\mathbf{\Theta}_g=\mathbf{\Theta}_g^0,\mathbf{\Theta}_d=\mathbf{\Theta}_d^0$
\For{$t=1, 2, \cdots$}
    \For{$v=1,2, \cdots, V$}
        \For{each $(\{\hat{\mathcal{I}}_m\}_{m=1}^{|\mathcal{M}_v|}, \mathbf{L}_v, \kappa^*)$}
            \State $\hat{\kappa}_v = \mathcal{G}(\mathbf{x}_v^{\text{LiDAR}}, \{\mathbf{x}_m^{\text{RGB}}\}_{m=1}^{|\mathcal{M}_v|})$;
            \State Compute $\nabla_{\mathbf{\Theta}_g^t} \mathcal{L}_{\text{handoff}}$ per Eq.~\eqref{eq10};
            \State Update $\mathbf{\Theta}_g^t=\mathbf{\Theta}_g^{t-1} - \eta_t\nabla_{\mathbf{\Theta}_g^t} \mathcal{L}_{\text{handoff}}$;
        \EndFor
    \EndFor
\EndFor
\For{$j=1, 2, \cdots$}
    \For{$m=1, 2, \cdots |\mathcal{M}_v|$}
        \For{each $(\hat{\mathcal{I}}_m, \mathbf{d}_m)$}
        \State $\hat{\mathbf{x}}_m^{\text{RGB}} = \text{UFE}(\hat{\mathcal{I}}_m)$;
        \State $\hat{\mathbf{d}}_m=\mathcal{D}(\hat{\mathbf{x}}_m^{\text{RGB}})$;
        \State Compute $\nabla_{\mathbf{\Theta}_m^j} \mathcal{L}_{\text{insp}}=\frac{\partial \mathcal{L}_{\text{insp}}}{\partial \hat{\mathbf{x}}_m^{\text{RGB}}}\frac{\partial \hat{\mathbf{x}}_m^{\text{RGB}}}{\mathbf{\Theta}_m^j}$ per Eq.~\eqref{eq11};
        \State Update $\mathbf{\Theta}_m^j=\mathbf{\Theta}_m^{j-1} - \eta_j\nabla_{\mathbf{\Theta}_m^j} \mathcal{L}_{\text{insp}}$;
        \EndFor
    \EndFor
\EndFor
\end{algorithmic}
\end{algorithm}

\section{Simulations}

\subsection{Experiment Setting}
\subsubsection{\textit{Dataset}} 
\begin{figure}[!t]
\centering
\includegraphics[width=\columnwidth]{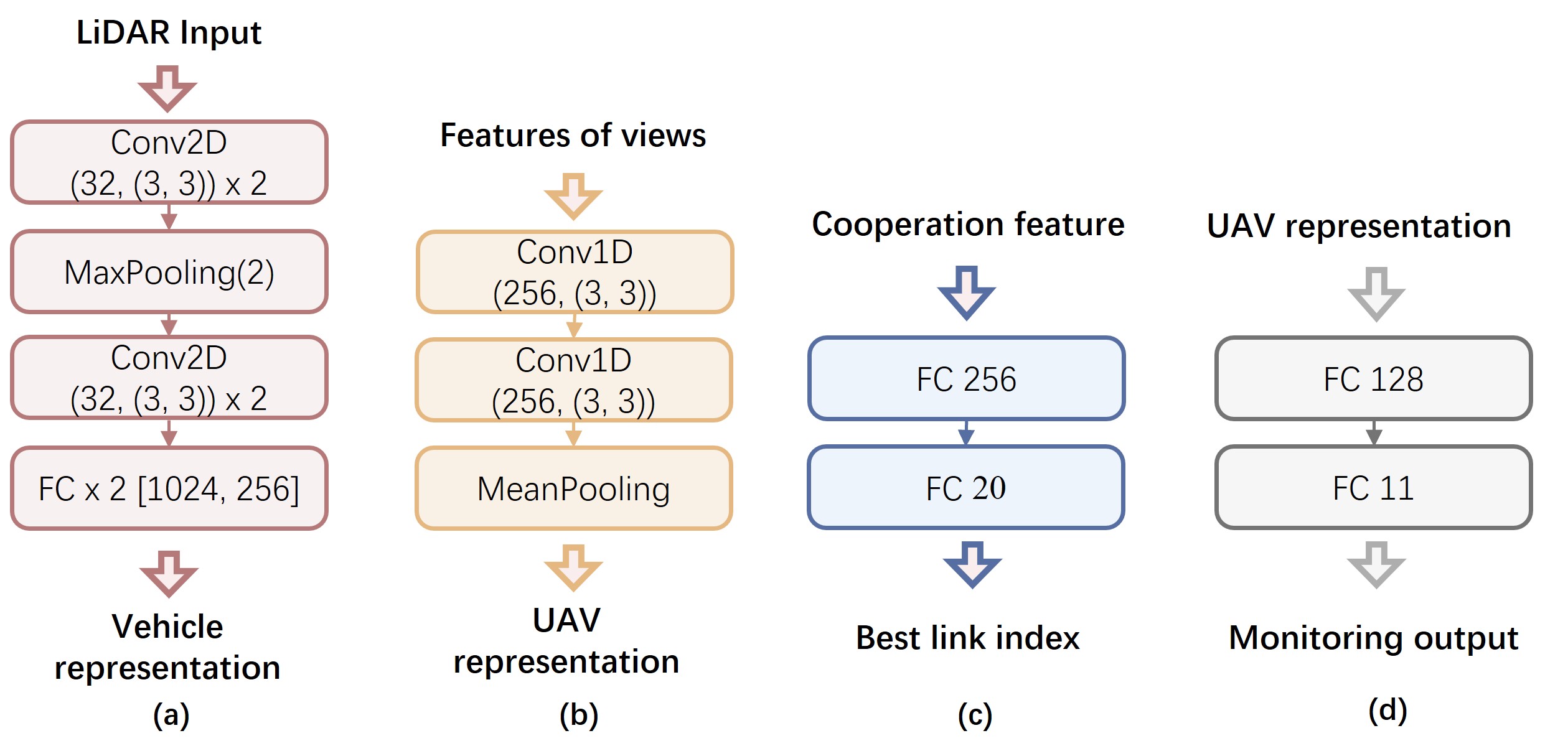}
\caption{Illustration of NN architectures for: (a) Designed CNN in VFE; (b) Designed CNN in UFE; (c) Designed FC for handoff; (d) TIH.}
\label{fig6}
\end{figure}

To demonstrate the capability of heterogeneous modalities cooperation in air-ground networks, we adopt the low-altitude economy scenario from the M\textsuperscript{3}SC dataset \cite{16}, which features packed buildings and non-line-of-sight (nLoS) communication conditions. The scenario consists of 4 RSUs and 6 UAVs equipped with $N_r=128$ and $N_u=32$ antennas, respectively. Each vehicle and each UAV are respectively equipped with LiDAR and RGB cameras, with a sampling frequency of 20 Hz. There is $10\%$ data missing to mimic the potential sensor failures. The downlink carrier frequency is 28 $\text{GHz}$. Among the 4 RSUs, there are 4 UAVs with qualified link are selected as relays. Thus, the number of connections is $|\mathcal{K}_v|=20$. The architectures of NNs customized for different sensing modalities and tasks are depicted in Fig.~\ref{fig6}.

\subsubsection{\textit{Implementation Details}} Table~\ref{tab1} summarizes the hyperparameter configurations used for training UAP-Net and TIH. The dataset is divided into three subsets: 72\% for training, 18\% for validation, and the remaining 10\% for testing. The training process is conducted on a workstation equipped with an NVIDIA RTX 3090 GPU. To mitigate memory consumption and ensure stable training, UAP-Net is trained using a gradient accumulation strategy with a step size of 8.

\begin{table}[!t]
    \centering
    \caption{\textsc{Hyper-parameters For Training}}
    \vspace{4pt}
    \label{tab1}
    \renewcommand{\arraystretch}{1.1}
    \begin{tabular}{c|c}
        \toprule[0.35mm]
        \textbf{Parameter} & \textbf{Value}~[UAP-Net, TIH] \\
        \midrule[0.15mm]
        Batch size & [64, 1024]\\
        Epochs & [40, 200] \\
        Optimizer & AdamW \\
        Learning rate & [$2\times10^{-4}$, $2\times10^{-3}$] \\
        \bottomrule[0.35mm]
    \end{tabular}
    \vspace{-8pt}
\end{table}

\subsubsection{\textit{Benchmarks}}
\begin{itemize}
    \item \textbf{Signal Measurements in Low-altitude Systems \cite{3}:}  
    Reactive handoff scheme with ground truth signal measurements in low-altitude systems is used as an ideal benchmark. It is the upper bound of the performance.

    \item \textbf{Signal Measurements in Ground Network \cite{3}:}  
    Reactive handoff scheme with ground truth signal measurements in ground network, so only 4 DLs are available.

    \item \textbf{Handoffs Prediction Based on RGB Images:}
    Deep learning-based proactive handoff schemes predict the optimal connection in low-altitude systems using RGB images captured by UAVs.

    \item \textbf{Handoffs Prediction Based on LiDAR:}
    Deep learning-based proactive handoff schemes predict the optimal connection in ground network using LiDAR cloud points obtained by vehicle.

    \item \textbf{YOLO-based Inspection:}  
    The pretrained YOLO model \cite{17} is employed as a benchmark to assess traffic monitoring performance.
\end{itemize}

\subsection{Outage Performance}
\begin{figure}[!t]
\centering
\includegraphics[width=0.9\columnwidth]{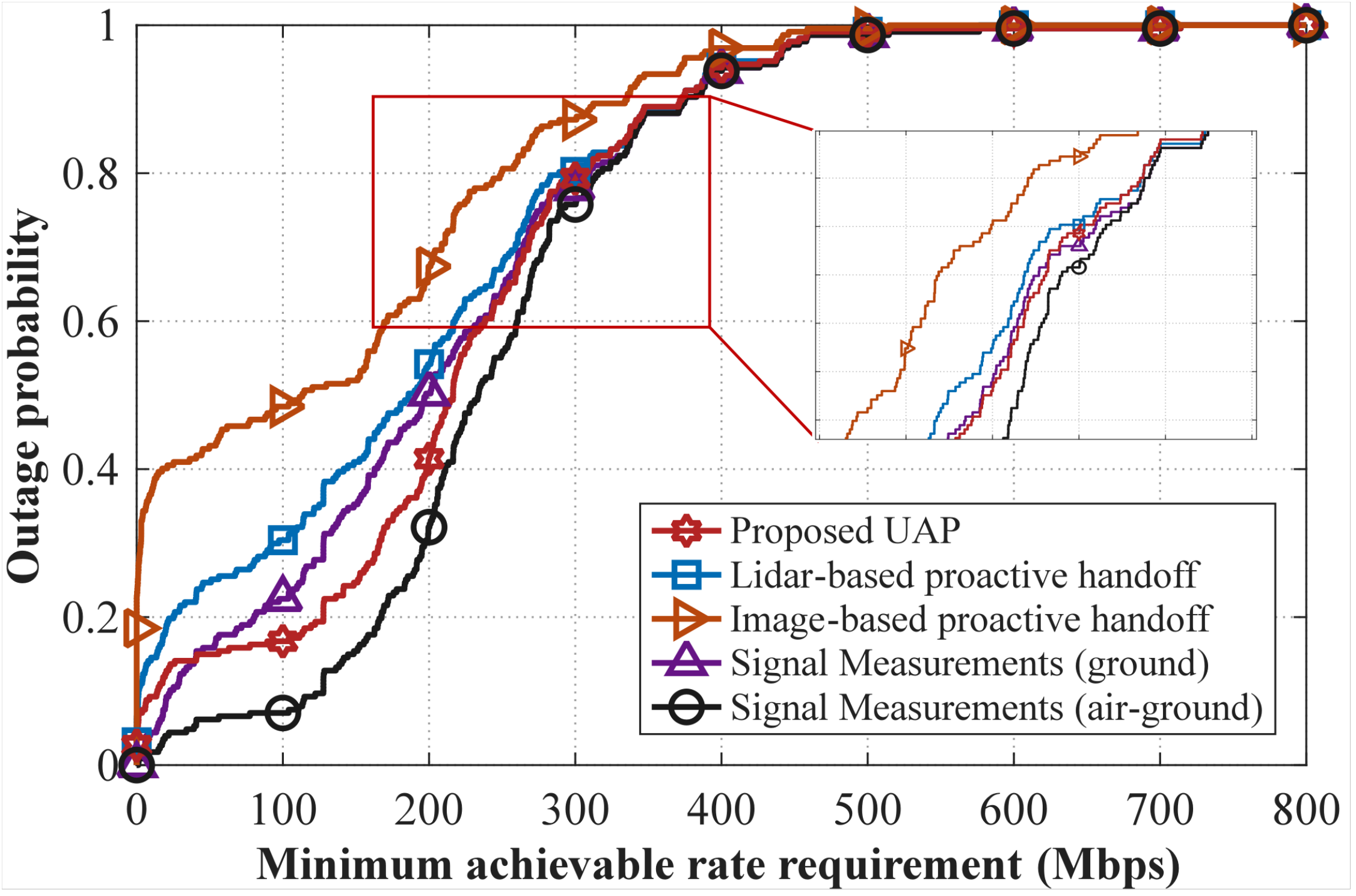}
\caption{Comparisons of outage probability performance among UAP and benchmarks.}
\label{fig3}
\end{figure}

To evaluate the performance of proposed UAP scheme on the system outage capacity, we analyze the outage probability under different handoff strategies. Let $P(R_T) = 1 - \frac{N(R_T)}{N_t}$ denote the outage probability, where $N(R_T) = \sum_{n=0}^{N} \chi(R_n, R_T)$ denotes the number of test samples whose achievable rate is higher than the minimum required achievable rate $R_T$. Here, $R_n$ represents the achievable rate at the $n$-th test sample from a total of $N_t$ samples. The indicator function is defined as:

\begin{equation}
\chi(R_n, R_T) =
\begin{cases}
1, & R_n \ge R_T, \\[4pt]
0, & \text{otherwise}.
\end{cases} \label{eq:16c}
\end{equation}
Fig.~\ref{fig3} illustrates that the proposed scheme achieves a lower outage probability than other proactive handoff schemes and signal measurement based method in ground networks when $R_T \leq 283~\text{Mbps}$. Specifically, when $R_T = 200~\text{Mbps}$, the UAP scheme reduces the outage probability by approximately 10\% compared with the method based on signal measurements in ground networks. When $R_T > 283~\text{Mbps}$, most optimal downlink connections are established directly with RSUs, resulting in comparable performance among all schemes.

\begin{figure}[!t]
\centering
\includegraphics[width=0.9\columnwidth]{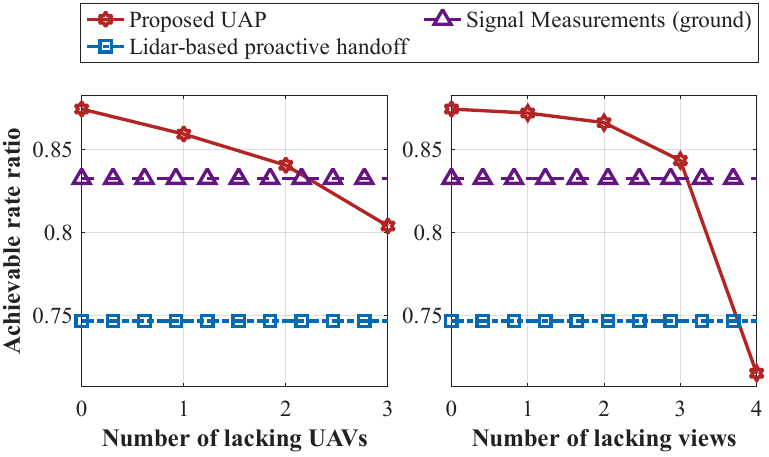}
\caption{Communication Performance of UAP in various involved views.}
\label{fig4}
\end{figure}

\subsection{Performance versus involvement of UAVs}

Fig.~\ref{fig4} shows the performance of UAP under under various views of involved UAVs, measured by the achievable rate ratio, defined as $\frac{\sum^{N_t}_{n=1} R_n}{\sum^{N_t}_{n=1} R_n^*}$, where $R_n^*$ is the performance upper bound. When the number of lacking UAVs increases to three, the achievable rate ratio of UAP only decreases less than 2\% compared with full UAV case, indicating that the ACAF effectively mitigates performance degradation. When each UAV lacks three views, the achievable rate ratio remains approximately 84\%, outperforming that of the measurement-based approach in ground networks. This confirms the benefit of global semantic information provided by UAVs. Additionally, the rate ratio becomes lower than that of the LiDAR-based proactive handoff method because the predefined 10\% data loss in the dataset causes some empty test samples. 

\begin{figure}[!t]
\centering
\includegraphics[width=0.9\columnwidth]{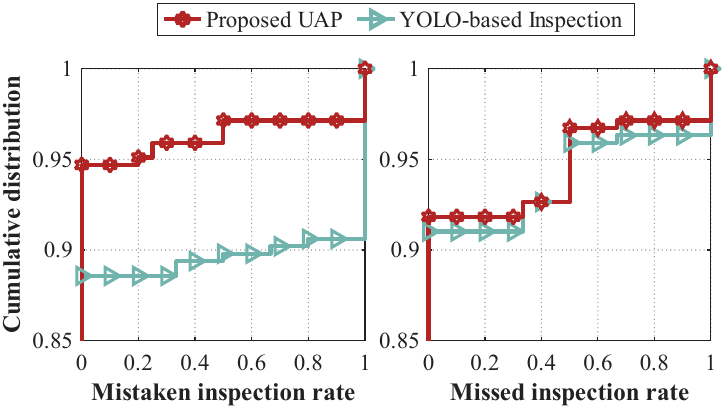}
\caption{Comparisons of inspection performance among UAP and YOLO.}
\label{fig5}
\vspace{-8pt}
\end{figure}


\subsection{Monitoring Performance}

Fig.~\ref{fig5} presents the traffic monitoring performance of UAVs when relaying signal. We analyze the cumulative distribution of the mistaken and missed inspection rate under different monitoring strategies. UAP achieves a comparable performance on missed inspection rate to YOLO-based method, while significantly increasing approximately 7\% cumulative distribution when minimum required mistaken inspection rate is 30\%, which is attributed to the UFE in UAP-Net, which focuses on extracted semantic features of vehicles and effectively suppresses the interference of irrelevant semantics. This results the benefit of global semantic information provided by UAVs. This successfully demonstrates the effectiveness of the dual-use UAV architecture.




\section{Conclusions}

This paper presented a dual-use UAV strategy that unified communication and monitoring functions in low-altitude networks. By developing cooperative perception modules for proactive handoff and a shared inspection module for traffic monitoring, we enabled UAVs to simultaneously maintain communication reliability and perform sensing tasks. Our two-stage training and distributed execution scheme ensured efficient operation, with simulations confirming that this integrated approach achieved enhanced communication performance without compromising monitoring accuracy,  demonstrating the practical viability of dual-use UAV architectures.



\bibliographystyle{IEEEtran}
\bibliography{reference}

\end{document}